\documentclass[twoside]{dis07}
\usepackage[latin1]{inputenc}
\usepackage[dvips]{graphicx,epsfig,color}
\usepackage{wrapfig,rotating}
\usepackage{amssymb,amsmath,array}
\usepackage{bm}

\begin{document}
\title{Rapidity gap survival in the black--disk regime}
\author{L.~Frankfurt$^1$, C.E.~Hyde$^2$, M.~Strikman$^3$, C.~Weiss$^4$
\thanks{Notice: Authored by Jefferson Science Associates, LLC under 
U.S.\ DOE
Contract No.~DE-AC05-06OR23177. The U.S.\ Government retains a
non-exclusive, paid-up, irrevocable, world-wide license to publish or
reproduce this manuscript for U.S.\ Government purposes.}
\vspace{.3cm}\\
1- School of Physics and Astronomy, Tel Aviv University, Tel Aviv, Israel
\vspace{.1cm}\\
2- Old Dominion University, Norfolk, VA 23529, USA, \\
and Laboratoire de Physique Corpusculaire, Universit\'e Blaise Pascal, 
63177 Aubi\`ere, France
\vspace{.1cm}\\
3- Department of Physics, Pennsylvania State University,
University Park, PA 16802, USA
\vspace{.1cm}\\
4- Theory Center, Jefferson Lab, Newport News, VA 23606, USA
}

\maketitle

\begin{abstract}
We summarize how the approach to the black--disk regime (BDR)
of strong interactions at TeV energies influences rapidity gap survival
in exclusive hard diffraction $pp \rightarrow p + H + p \; 
(H = \textrm{dijet}, \bar Q Q, \textrm{Higgs})$.
Employing a recently developed partonic description of such processes,
we discuss (a) the suppression of diffraction 
at small impact parameters by soft spectator interactions in the BDR;
(b) further suppression by inelastic interactions of hard spectator 
partons in the BDR; (c) correlations between hard and soft interactions. 
Hard spectator interactions substantially reduce the 
rapidity gap survival probability at LHC energies compared 
to previously reported estimates.
\end{abstract}

\section{Introduction}
At high energies strong interactions enter a regime in which 
cross sections are comparable to the ``geometric size'' of 
the hadrons, and unitarity becomes an essential feature of 
the dynamics. By analogy with quantum--mechanical scattering
from a black disk, in which particles with impact
parameters $b < R_{\text{disk}}$ experience inelastic interactions 
with unit probability, this is known as the black--disk
regime (BDR). The approach to the BDR is well--known
in soft interactions, where it generally can be attributed to the 
``complexity'' of the hadronic wave functions. It is seen \textit{e.g.}\
in phenomenological parametrizations of the $pp$ elastic scattering 
amplitude, whose profile function $\Gamma(b)$ approaches unity at 
$b = 0$ for energies $\sqrt{s} \gtrsim 2 \, \textrm{TeV}$. More recently
it was realized that the BDR is attained also in hard processes 
described by QCD, due to the increase of the gluon density in the proton 
at small $x$. Theoretically, this phenomenon can be studied in the 
scattering of a small--size color dipole ($d \sim 1/Q$) from the proton. 
Numerical studies show that that at TeV energies the dipole--proton
interaction is close to ``black'' up to $Q^2 \sim 
\textrm{several 10 GeV}^2$ \cite{Frankfurt:2005mc}. This fact has 
numerous implications for the dynamics of $pp$ collisions at the LHC,
where multiple hard interactions are commonplace. For example, 
it predicts dramatic changes in the multiplicities and $p_T$ spectra 
of forward particles in central $pp$ collisions
compared to extrapolations of the Tevatron data \cite{Frankfurt:2003td}.
Absorption and energy loss of leading partons by inelastic interactions
in the BDR can also account for the pattern of forward pion production 
in $d$--$Au$ collisions at STAR \cite{Frankfurt:2007rn}.

Particularly interesting is the question what the approach to the 
BDR implies for exclusive hard diffractive scattering,
$pp \rightarrow p + H + p$. In such processes a high--mass system
($H = \textrm{dijet}, \bar QQ, \textrm{Higgs})$ is produced in
a hard process involving exchange of two gluons between the protons.
At the same time, the spectator systems must interact in a way such 
as not to produce additional particles. This restricts the set
of possible trajectories in configuration space and results in a 
suppression of the cross section compared to non-diffractive events. 
For soft spectator interactions this suppression is measured by the
so--called rapidity gap survival (RGS) probability. Important questions
are (a) what role the BDR plays in traditional soft--interaction RGS; 
(b) how the physical picture of RGS is modified by hard spectator 
interactions in the BDR at LHC energies; (c) how possible correlations
between hard and soft interactions affect RGS. 
These questions can be addressed in a recently proposed partonic
description of exclusive diffraction \cite{Frankfurt:2006jp}, 
based on Gribov's parton 
picture of high--energy hadron--hadron scattering. This approach 
allows for a model--independent treatment of the interplay of hard
and soft interactions, and for the consistent implementation of 
the BDR at high energies (for details, see Ref.~\cite{Frankfurt:2006jp}). 
\section{Black--disk regime in soft spectator interactions}
\begin{figure}
\begin{tabular}{lcl}
\parbox[c]{0.3\columnwidth}{
\includegraphics[width=0.3\columnwidth]{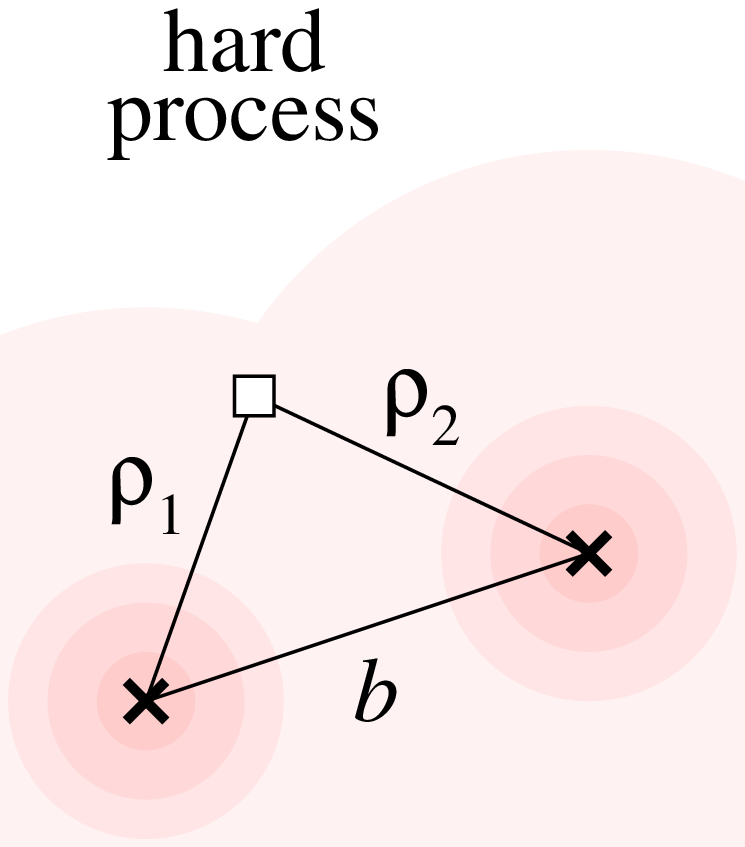}
} 
& \hspace{0.01\columnwidth} &
\parbox[c]{0.6\columnwidth}{
\includegraphics[width=0.6\columnwidth]{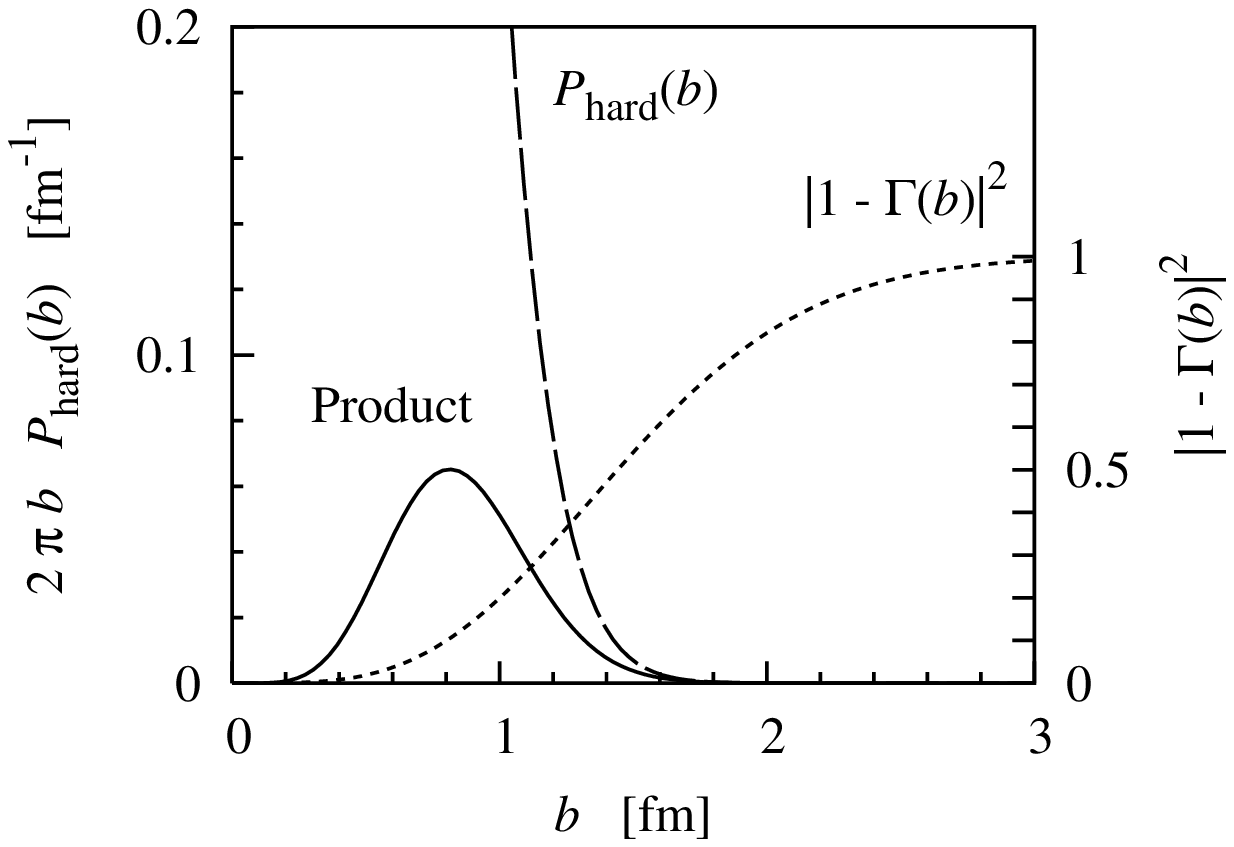}}
\\[-3ex]
(a) & & (b)
\end{tabular}
\caption{(a) Transverse geometry of hard diffractive $pp$ scattering.
(b) Dashed line: Probability for hard scattering process 
$P_{\text{hard}} (\bm{b})$ as function of the $pp$ impact parameter, $b$.
Dotted line: Probability for no inelastic interactions between
the protons, $|1 - \Gamma (\bm{b})|^2$.
Solid line: Product $P_{\text{hard}} (\bm{b}) |1 - \Gamma (\bm{b})|^2$.
The RGS probability (\ref{survb}) is given by the area under this curve.
The results shown are for Higgs production at the LHC
($\sqrt{s} = 14\, \textrm{TeV}, M_H \sim 100\, \textrm{GeV}$).}
\label{Fig:rgs}
\end{figure}

A simple ``geometric'' picture of RGS is obtained in the approach 
of Ref.~\cite{Frankfurt:2006jp} in the approximation where hard and 
soft interactions are considered to be independent. The hard two--gluon
exchange process can be regarded as happening locally in space--time 
on the typical scale of soft interactions. In the impact parameter
representation (see Fig.~\ref{Fig:rgs}a) the RGS probability can be
expressed as
%
%
\begin{equation}
S^2 \;\; = \;\; 
\int d^2 b \; P_{\text{hard}} (\bm{b}) \; |1 - \Gamma (\bm{b})|^2 .
\label{survb}
\end{equation}
Here $P_{\text{hard}} (\bm{b})$ is the probability for two hard gluons 
from the protons to collide in the same space--time point, given by
the overlap integral of the squared transverse spatial distributions
of gluons in the colliding protons, normalized to 
$\int d^2 b \; P_{\text{hard}} (\bm{b}) = 1$
(see Fig.~\ref{Fig:rgs}b). 
The function $|1 - \Gamma (\bm{b})|^2$ is the probability for the
two protons not to interact inelastically in a collision with
impact parameter $b$. The approach to the BDR in $pp$ scattering
at energies $\sqrt{s} \gtrsim 2\, \textrm{TeV}$
implies that this probability is practically zero at small 
impact parameters, and becomes significant only for 
$b \gtrsim 1\, \textrm{fm}$ (see Fig.~\ref{Fig:rgs}b).
This eliminates the contribution from small
impact parameters in the integral (\ref{survb}) 
(see Fig.~\ref{Fig:rgs}b) and determines the value of the 
RGS probability to be $S^2 \ll 1$.
It is seen that the approach to the BDR in soft interactions
plays an essential role in the physical mechanism of RGS.
\section{Black--disk regime in hard spectator interactions}
%
%
\begin{figure}
\begin{tabular}{lcl}
\parbox[c]{0.25\columnwidth}{
\includegraphics[width=0.25\columnwidth]{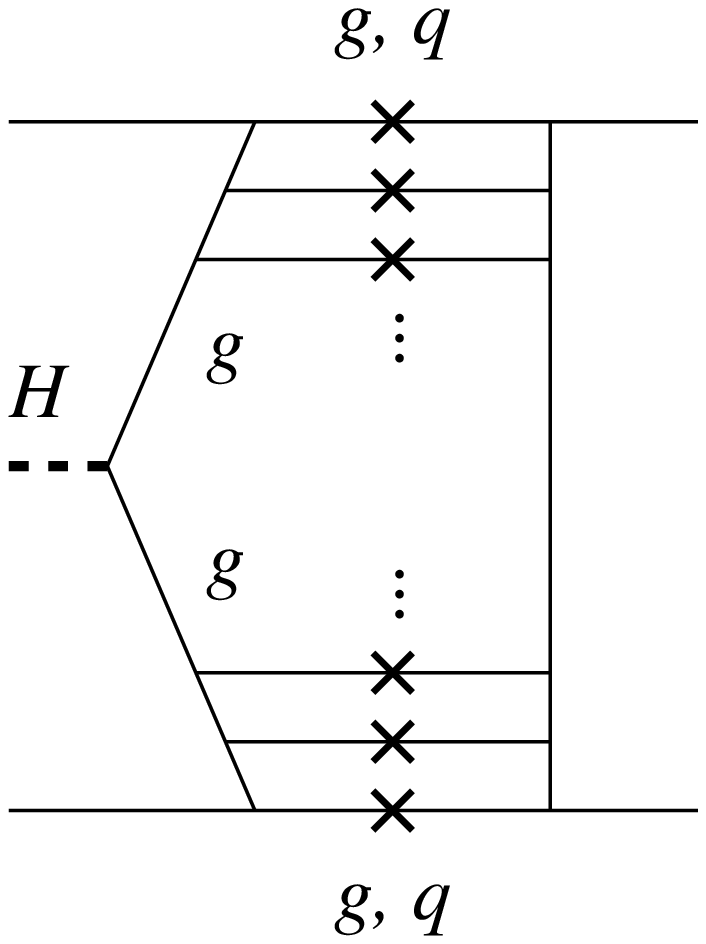}}
&
\hspace{.15\columnwidth}
&
\parbox[c]{0.45\columnwidth}{
\includegraphics[width=0.45\columnwidth]{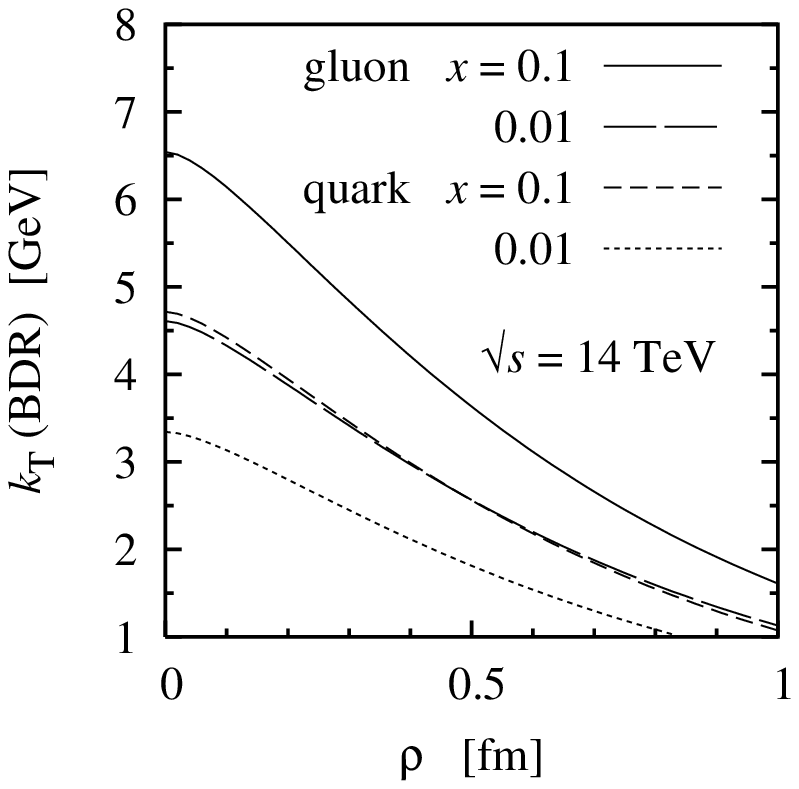}}
\\[-3ex]
(a) & & (b)
\end{tabular}
\caption{(a) Absorption of parent partons in the evolution
by interactions in the BDR. (b) The critical 
transverse momentum, $k_T(\textrm{BDR})$, 
below which partons are absorbed with high probability
($|\Gamma^{\textrm{parton-proton}}| > 0.5$), as a function of the
parton--proton impact parameter, $\rho$.}
\label{Fig:hardscreen}
\end{figure}
At LHC energies even highly virtual partons ($k^2 \sim \textrm{few GeV}^2$)
with $x \gtrsim 10^{-2}$ experience ``black'' interactions with the 
small--$x$ gluons in the other proton. This new effect causes
an additional suppression of diffractive scattering which is not 
included in the traditional RGS probability \cite{Frankfurt:2006jp}. 
One mechanism by which this happens is the absorption of ``parent''
partons in the QCD evolution leading up to the
hard scattering process (see Fig.~\ref{Fig:hardscreen}a). 
Specifically, in Higgs production at the LHC the gluons producing 
the Higgs have momentum fractions 
$x_{1, 2} \sim M_H / \sqrt{s} \sim 10^{-2}$; their ``parent'' partons
in the evolution (quarks and gluons)
typically have momentum fractions of the order 
$x \sim 10^{-1}$ and transverse momenta $k_T^2 \sim \textrm{few GeV}^2$.
Quantitative studies of the BDR in the dipole picture show
that at the LHC energy such partons are absorbed with near--unit 
probability if their impact parameters with the other proton 
are $\rho_{1, 2} \lesssim 1 \, \textrm{fm}$ 
(see Fig.~\ref{Fig:hardscreen}b). For proton--proton impact parameters 
$b < 1 \, \textrm{fm}$ about $90\%$ of the strength in 
$P_{\textrm{hard}}(b)$ comes from parton--proton impact parameters 
$\rho_{1, 2} < 1 \, \textrm{fm}$ (\textit{cf}.\ Fig.~\ref{Fig:rgs}a),
so that this effect practically eliminates diffraction at 
$b < 1 \, \textrm{fm}$. Since $b < 1 \, \textrm{fm}$ accounts for 2/3 
of the cross section [see Eq.~(\ref{survb}) and Fig.~\ref{Fig:rgs}b)],
and the remaining contributions at $b > 1 \, \textrm{fm}$ are also 
reduced by absorption, we estimate that inelastic interactions
of hard spectators in the BDR reduce the RGS probability at LHC
energies to about 20\% of its soft--interaction value. 
(Trajectories with no emissions,
corresponding to the $\delta(1 - x)$ term in the evolution kernel,
are not affected by absorption; however, their contributions
are small because they effectively probe the gluon density at 
the soft input scale.) Since this effect ``pushes'' diffractive $pp$ 
scattering to even larger impact parameters than allowed by 
soft--interaction RGS it should also manifest itself in a shift of the 
final--state proton transverse momentum distribution to smaller values, 
which could be observed in $p_T$--dependent measurements 
of diffraction at the LHC.

The estimates reported here are based on the assumption that DGLAP 
evolution reasonably well describes the gluon density down to 
$x \sim 10^{-6}$; the quantitative details (but not the basic picture) 
may change if small--$x$ resummation corrections were to significantly 
modify the gluon density at such values of $x$ 
(see Ref.~\cite{Ciafaloni:2007gf} and references therein). 
The effect of hard spectator interactions described here
is substantially weaker at the Tevatron energy.
\section{Correlations between hard and soft interactions}
%
%
\begin{wrapfigure}{l}{0.45\columnwidth}
\centerline{\includegraphics[width=0.4\columnwidth]{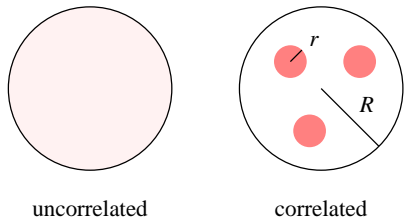}}
\caption{Transverse parton correlations.}
\label{Fig:corr}
\end{wrapfigure}
The partonic approach to RGS of Ref.~\cite{Frankfurt:2006jp} also
allows one to incorporate effects of correlations in the partonic
wavefunction of the protons. They can lead to correlations between
hard and soft interactions in diffraction, which substantially
modify the picture of RGS compared to the independent interaction
approximation. The CDF data on $pp$ collisions with multiple hard
processes indicate the presence of substantial transverse 
correlations of size $r \ll R$ between partons (see Fig.~\ref{Fig:corr}). 
A possible explanation of their origin could be 
the ``constituent quark'' structure of the proton suggested by
the instanton vacuum model of chiral symmetry breaking.
Such correlations modify the picture of RGS in hard diffractive 
$pp$ scattering compared to the independent interaction 
approximation in two ways \cite{Frankfurt:2006jp}. 
On one hand, with correlations inelastic interactions between
spectators are much more likely in configurations in which two 
large--$x$ partons collide in a hard process than in average 
configurations, reducing the RGS probability compared to the
uncorrelated case. On the other hand, the ``lumpiness'' implies
that there is generally a higher chance for the remaining spectator 
system not to interact inelastically compared to the mean--field
approximation. A quantitative treatment of correlations in RGS,
incorporating both effects, remains an outstanding problem.

%
%
\begin{footnotesize}
%


\end{footnotesize}

\end{document}